\documentclass[12pt]{article}
\usepackage{amsthm,amsmath,amssymb}
\usepackage[utf8]{inputenc}
\usepackage{lineno,hyperref}
\usepackage{bm}
\usepackage{xcolor}
\usepackage{amssymb}
\usepackage{amsmath}
\usepackage{mathrsfs}
\usepackage{rotating}
\usepackage{geometry}
\usepackage[ruled,linesnumbered]{algorithm2e}
\usepackage{setspace}
\usepackage{comment}
\usepackage{tabularx}
\usepackage{longtable}
\usepackage{booktabs}
\usepackage[center]{subfigure}
\usepackage{captcont}

\newtheorem{theorem}{Theorem}
\newtheorem{lemma}{Lemma}
\newtheorem{remark}{Remark}

\def\st{\mbox{s.~t.}}

\newcommand{\ceil}[1]{\left\lceil #1 \right\rceil}
\newcommand{\balpha}{\boldsymbol \alpha}
\newcommand{\bbeta}{\boldsymbol \beta}

\usepackage{natbib}
\usepackage{graphicx}

\newcommand{\blind}{0}
\begin{document}

\def\spacingset#1{\renewcommand{\baselinestretch}%
{#1}\small\normalsize} \spacingset{1}
\if0\blind
{
  \title{\bf Change-Point Detection in Time Series \\ Using Mixed Integer Programming
  }
  
 \author{Artem Prokhorov\\
   University of Sydney, CEBA, CIREQ\\
   \and 
   Peter Radchenko \\
   University of Sydney \\
   \and 
   Alexander Semenov \\
   University of Florida \\
   \and 
   Anton Skrobotov \\
   HSE University}
  \maketitle
} \fi
\if1\blind
{
    \title{\LARGE\bf Change-Point Detection in Time Series \\ Using Mixed Integer Programming}
\maketitle
  \medskip
} \fi
\bigskip
\begin{abstract}
We use cutting-edge mixed integer optimization (MIO) methods to develop a framework for detection and estimation of structural breaks in time series regression models. The framework is constructed based on the least squares problem subject to a penalty on the number of breakpoints. We restate the $l_0$-penalized regression problem as a quadratic programming problem with integer- and real-valued arguments and show that MIO is capable of finding provably optimal solutions using a well-known optimization solver. Compared to the popular $l_1$-penalized regression (LASSO) and other classical methods, the MIO framework permits simultaneous estimation of the number and location of structural breaks as well as regression coefficients, while accommodating the option of specifying a given or minimal number of breaks. We derive the asymptotic properties of the estimator and demonstrate its effectiveness through extensive numerical experiments, confirming a more accurate estimation of multiple breaks as compared to popular non-MIO alternatives. Two 
empirical examples demonstrate usefulness of the framework in applications from business and economic statistics. 
\end{abstract}

\noindent%
\vfill
\newpage
\spacingset{1.8} 

\section{Introduction}

Identifying structural breaks in time series, also known as change-points, regime shifts and concept drifts, is a major area of interest within theoretical and applied statistics, going back at least to the 1960s \cite[see, e.g.,][]{Shiryaev, Roberts}. In modern econometrics the focus has been on statistical approaches that estimate breakpoints by minimizing the regression sum of squares \cite[see, e.g.,][]{bai1998estimating,  bai2003computation} or $l_1$-penalized sum of squares \cite[see, e.g.,][]{qian_su_2016, kaddoura/westerlund:23}. As one of the most highly-cited examples, \cite{bai1998estimating} proposed a specific-to-general testing strategy for estimating the number of breaks in linear regression models with potential heterogeneity in the errors. The method requires testing the null hypotheses of $m$ breaks against the alternative of $m+1$ break starting with $m=0$. The estimated number of breaks then is that for which the null hypothesis is not rejected. 

Testing-based approaches have been criticized for not always offering a consistent estimator of break dates and for the tendency to overestimate the true number of breaks with a positive probability, equal to the tests' significance level asymptotically. \cite{bai2003computation} suggested using information criteria to choose the number of breaks, providing a consistent estimator of the break number. However, the approaches inevitably put  
restrictive assumptions on the minimal length of a regime to be set by the researcher, while the critical values crucially depend on this length both in large and in small samples. 

Penalized methods have been proposed to circumvent the  restriction on the minimal length of regimes. The LASSO (Least Absolute Shrinkage and Selection Operator) of \citet{tibshirani1996regression} has been extremely effective at 
selecting the number of regression parameters with a simultaneous estimation of the non-zero parameters in linear models. \cite{harchaoui2010multiple} and \cite{bleakley2011group} consider the estimation of  break locations in one-dimensional piece-wise constant signals, under the assumption of independence. \citet{chan2014group} extend their approach to dependent data allowing the number of breaks to grow with the sample size. They also provide a justification for using a second step in the selection procedure in order to prevent an overestimation of the number of breaks. 

\citet{behrendt2021note} proposed using adaptive group LASSO to select the number of breaks consistently as an alternative to the two-step procedure of \citet{chan2014group}. The two-step procedure of \citet{chan2014group} is easier to apply, but it is less efficient than adaptive group LASSO of \citet{behrendt2021note}. \cite{qian_su_2016} also considered a linear regression model and estimated the number of regimes and model parameters by using adaptive fused LASSO. Their approach is also two-step due to an overestimation of the true break date in the first step.


In the context of these developments, it has been a common belief that mixed integer optimization (MIO) is not suitable for such problems due to what is known as ``combinatorial explosion'', that is, the explosive growth in the number of combinations to consider and associated insurmountable computational task. However,  recent remarkable advances in computational and algorithmic methods of optimization over integer-valued arguments have shown attractive properties of integer and mixed integer programming as a means of obtaining efficient and provably optimal solutions in a wide range of statistical problems \cite[see, e.g.,][]{bertsimas2016, mazumder2023subset, hazimeh2023grouped, gomez2021mixed, rebennack2020piecewise}.

For example, the problem of subset selection has become feasible and even standard in applications with datasets much larger than statisticians previously thought possible \cite[see, e.g.,][]{bertsimas/etal:20}. Moreover, it is often suggested that subset selection using integer programming outperforms LASSO in many situations. The MIO challenge was famously picked by statisticians in a  recent issue of Statistical Science \cite[see, e.g.,][and rejoinders therein]{hastie/etal:20}. Such developments raise the prospect of applying MIO in other settings of interest to econometricians. For example, \cite{lee/etal:17} propose and implement MIO for a change-point regression model with one change and apply it to model the US unemployment rate.\footnote{We thank an anonymous referee for pointing out this paper to us. The paper considers a similar model to ours, but with only two regimes and the difference in the slope parameters determined by possibly unobserved factors. It also designs advanced computational methods for when the dimension of $\beta$ grows. In contrast, our approach focuses on the estimation of constant slope coefficients with an unrestricted number and location of breaks.}   

In this paper we develop an MIO-based framework for simultaneously  estimating the number and  location of structural breaks as well as the parameters of a time series regression model. Formulated as a mixed integer quadratic programming problem, the new method shows an attractive performance compared to LASSO-type procedures, especially when the number of breaks is high. Existing solvers quickly obtain solutions of the problem to optimality, and no multi-step procedures, typical for LASSO-based competitors, are necessary. Importantly, our framework permits estimation of the unknown number of structural breaks while  accommodating the option of specifying a required or minimal number of breaks if necessary. 
As we show, the new estimator enjoys relative simplicity and attractive theoretical properties under assumptions that are somewhat weaker than in the alternatives available in the literature. 

To demonstrate the effectiveness of our approach, we conduct comprehensive Monte Carlo simulations, using a well-known and easily accessible solver. 
We show how to choose the tuning parameters and compare our results with established methods such as those proposed by \cite{bai1998estimating} and \cite{qian_su_2016}. 

The paper is organized as follows. Section \ref{sec:model} formulates the model and assumptions, and proposes our MIO approach. In Section \ref{sec:theory}, we state the additional assumptions and main asymptotic results on consistency and asymptotic normality of the proposed estimators. Numerical experiments are discussed in Section \ref{sec:sims}. Empirical applications are provided in Section \ref{sec:empirics}. Section \ref{sec:conclude} concludes. All proofs and technical details are collected in the Appendix.

\section{Methodology}\label{sec:model}


We assume that the data is generated by the following process:
\begin{equation}\label{eq:regression}
y_t = {\bbeta_t^*}^{\top} x_t + u_t,
\end{equation} 
where $x_t$ is a $p\times 1$ vector of regressors, $u_t$ is the error term, and the $(p\times 1)$ vector $\bbeta^*_t$ takes  distinct vector values $\balpha^*_j, j=1, \ldots, m^*+1$, in the time interval $T^*_{j-1} \le t < T^*_{j}$, where $m^*$ is the number of breakpoints and where we use the convention that $T^*_{0}= 1$ and $T^*_{m^*+1}= T+1$. In this model, the indices $(T^*_1,\dots,T^*_{m^*})$, or breakpoints, are assumed to be unknown.


The goal is to find the unknown number $m^*$ of unknown break dates $(T^*_1,\dots,T^*_{m^*})$ as well as the regression coefficients  $\balpha=({\balpha^*_1}^{\top},\dots, {\balpha^*_{m^*+1}}^{\top})^{\top}$. It is clear that with no penalty, an in-sample prediction error minimization for (\ref{eq:regression}) gives $\hat{m}=T-1$ breaks and a perfect fit. This solution is unlikely to generalize well out-of-sample. To avoid overfitting, it is natural to impose a penalty that counteracts the reduction in prediction error 
for adjacent values of $\bbeta^*_t$ that are not too far from one another. A common way of doing this is to utilize various forms of $l_1$-norm of the difference $\bbeta^*_t-\bbeta^*_{t-1}$; see, e.g., Group Fused LASSO (GFL) of \cite{qian_su_2016}, Grouped LASSO of \cite{kaddoura/westerlund:23}.

\begin{remark}
{\rm  Model \eqref{eq:regression} can be considered as pure structural change model in the \cite{bai1998estimating} terminology. At the same time, we can consider the so-called partial structural change model where some elements of $\bbeta_t$ do not sustain structural changes. In others words, $\bbeta_t$ can be decomposed into 
a $(p_1+p_2)$-vector $\bbeta_t=(\bbeta_{1t}^{\top}, \bbeta_{2t}^{\top})^{\top}$ with a $p_2$-subvector $\bbeta_{2t}=\bbeta_2$
 which does not depend on $t$. The pure structural change model is considered for exposition purpose and brevity, but all results for the pure structural change model can be extended to the partial structural change model with more tedious proofs. }
\end{remark}

The class of estimators we consider can be stated as the following $l_0$-penalized optimization problem:
\begin{equation}\label{l0-penal}
     \widehat{\bbeta}=\arg\min_{\bbeta} \sum_{t=1}^{T}(y_t - \bbeta_t^{\top} x_t)^2 +  \lambda\sum_{t=2}^{T}{\mathbf{1}\{\bbeta_t\ne \bbeta_{t-1}\}},
 \end{equation}
where $\mathbf{1}\{\cdot\}$ denotes the indicator function and $\lambda=\lambda_T$ is a positive tuning parameter. We note that our penalty function, $\sum_{t=2}^{T}{\mathbf{1}\{\bbeta_t\ne \bbeta_{t-1}\}}$, simply counts the number of jumps in the regression coefficient vectors $\bbeta_t$.  

It is interesting to compare and contrast our approach with the GFL method of \cite{qian_su_2016}, which is based on the following optimization problem:
\begin{equation*}
     \min_{\bbeta} \sum_{t=1}^{T}(y_t - \bbeta_t^{\top} x_t)^2 +  \lambda\sum_{t=2}^{T}\|\bbeta_t- \bbeta_{t-1}\|.
 \end{equation*}
The difference between the two estimators is in the penalty function. While our approach penalizes the number of jumps in the $\bbeta$-coefficients \textit{directly}, the GFL method does so \textit{indirectly}, by using a group LASSO-type penalty $\sum_{t=2}^{T}\|\bbeta_t- \bbeta_{t-1}\|$. This penalty serves as a proxy for our $\ell_0$ penalty $\sum_{t=2}^{T}{\mathbf{1}\{\bbeta_t\ne \bbeta_{t-1}\}}$, which counts the exact number of coefficient jumps. We note that one of the consequences of using a LASSO-type ($\ell_1$) penalty rather than an $\ell_0$ penalty is the resulting shrinkage of the estimated coefficient differences $\bbeta_t- \bbeta_{t-1}$. 

To leverage impressive recent advances in the field of mixed integer optimization (MIO), we propose solving~(\ref{l0-penal}) by formulating it as a MIO problem as follows:
\begin{subequations}
\begin{align} 
\min_{\bbeta_t, z_t} \quad & \sum_{t=1}^{T}(y_t - \bbeta_t^{\top} x_t)^2 +  \lambda\sum_{t=1}^{T-1}z_t,\label{est.eq1}\\
\st \quad & -M\boldsymbol{e}z_{t} \leq \bbeta_{t+1} - \bbeta_{t} \leq  M\boldsymbol{e}z_{t},   \ \ \ \text{for all } \, t=1, \dots, T-1, \label{est.eq2}\\
\quad &  z_t + z_{t+1} \leq 1   \ \ \ \text{for all }  \, t=1, \dots, T-2, \\
\quad & z_{t} \in \{0,1\}  \ \ \ \text{for all }  \, t=1, \dots, T-1,\label{est.eq4}
\end{align}
\end{subequations} 
where $\boldsymbol{e}$ is a $p$-vector of ones. Appendix A contains the details of why the two formulations are equivalent. 
Formulation (\ref{est.eq1})-(\ref{est.eq4}) involves optimization over continuous variables~$\bbeta$ and binary variables~$z_t$, where $z_t$ equals 1 if and only if there is a break 
at time $t$. In~(\ref{est.eq2}), $M$ is some very large pre-specified  constant, which gives name to the ``Big-M'' term for the general formulation of the above problem. We note that Big-M formulations are extensively used in mixed integer programming. In particular, they have recently been used in a number of~$\ell_0$-regularized regression approaches \citep[see, for example,][]{bertsimas2016,mazumder/radchenko:17,hazimeh2023grouped}. Optimization problem (\ref{est.eq1})-(\ref{est.eq4}) is tractable and can be efficiently solved in practice using state-of-the-art MIO solvers.

\begin{remark}
{\rm  
Constraint (3d) defines binary variables~$z_t$, which we use to control whether $\bbeta_{t+1} - \bbeta_{t}$ is zero or not. Constraints (3b)-(3c) have either computational or conceptual meaning, or both. Constraint (3b) is conceptual: it ensures that $\bbeta_t$ does not change in between breaks; when a break occurs, each element of $\bbeta_t$ can jump by no more than $M$, where $M$ is a large constant, serving a computational purpose.
Constraint (3c) serves the following computational purpose: it prevents the breaks from happening consecutively.
We note that we can generalize this constraint to $ z_t + z_{t+1} + z_{t+2}\leq 1$ or any other number of leads to force intervals to be at least that number of time periods long.  
}
\end{remark}

\section{Asymptotic properties}\label{sec:theory}

In this section we study the asymptotic properties of our proposed estimator~(\ref{l0-penal}). We start by introducing some notation. Let $I^*_j=T^*_j-T^*_{j-1}$ for $j=1,...,m^*+1$ and define
$$
I_{\text{min}}=\min_{1\le j\le m^*+1}I^*_j, \quad J_{\text{min}}=\min_{1\le j\le m^*}\|\balpha^*_{j+1}-\balpha^*_{j}\|, \quad 
\text{and} \quad J_{\text{max}}=\max_{1\le j\le m^*}\|\balpha^*_{j+1}-\balpha^*_{j}\|.
$$
We note that $I_{\text{min}}$ is the smallest interval length among the~$m^*+1$ regimes of the true data-generating process, while $J_{\text{min}}$ and $J_{\text{max}}$ measure the smallest and largest jump sizes, respectively, in the true vector of coefficients.

The main result in this section establishes consistency of our approach in estimating the true number of breaks, breakpoints, and regression coefficients and also derives the corresponding rates of convergence. 
This result corresponds to the combination of the following two theorems in \cite{qian_su_2016}: Theorem 3.4 (on correctly estimating the true number of breaks using the information criterion) and Theorem 3.1 (on the rate of convergence for the breakpoints and coefficients when the correct number of breaks is used). We impose the same assumptions (A1 and A2) on the $\{(x_t,u_t)\}$ process as \cite{qian_su_2016} do in their theoretical analysis. In particular, we assume that $\{(x_t,u_t)\}$ is a strong mixing process with a geometric decay rate and impose moment conditions on the corresponding random variables. We also bound the eigenvalues of $\frac1{r-s}\sum_{t=s}^{r-1} E(x_tx_t^{\top})$ for $r-s\ge T\delta_T$  and lower-bound~$\delta_T$ depending on whether the finite-moment or the exponential-moment condition is satisfied. More specifically, if $(x_t,u_t)$ have uniformly bounded $4q$-th moments for some $q>1$, then $\delta_T$ cannot go to zero faster than $T^{(1-q)/q}$; alternatively, if $Ee^{c(\|x_t\|+|u_t|)^{2\gamma}}$ are uniformly bounded for some positive $c,\gamma$, then the fastest rate of decrease for~$\delta_T$ is $(\log T)^{(2+\gamma)/\gamma}/T$.
Assumptions A1 and A3 are formally stated in Appendix B. We also impose the following additional requirements.

\noindent\textbf{Assumption A3}.
\begin{enumerate}
\item[(i)] $J_{\textrm{max}}=O\big(1\big)$ and $T\delta_T J^2_{\textrm{min}}/(\log T)^{c_{\delta}}\rightarrow\infty$ as $T\rightarrow\infty$, where $c_{\delta}=6$ if A1(ii.a) is satisfied and $c_{\delta}=1$ if A1(ii.b) is satisfied.
\item[(ii)] $\delta_T=O\big(I_{\textrm{min}}^{1/2}/T\big)$ and $T^{1/2}m^*\big( I_{\textrm{min}} J^2_{\textrm{min}} \big)^{-1}\rightarrow0$ as $T\rightarrow\infty$.
\end{enumerate}
This assumption is a weaker version of Assumption A3 in \cite{qian_su_2016}, which is required for their Theorem 3.4. More specifically, we do not impose their conditions $Tm^*\big[(\log I_{\textrm{min}})^{c_{\delta}/2}T^{-1/2}I_{\textrm{min}}^{-1/2}+I_{\textrm{min}}^{-1} \big]\big( I_{\textrm{min}} J^2_{\textrm{min}} \big)^{-1}\rightarrow0$ and $m^*=O\big(\log T\big)$.

We define $\widehat{m}$, $\widehat{\balpha}_j$, $\widehat{T_j}$ as the characteristics of our estimator (\ref{l0-penal}) that are the analogs of the corresponding population quantities ${m}^*$, ${\balpha}^*_j$,  and~${T}^*_j$.
\begin{theorem}
\label{asympt.theorem}
Suppose that Assumptions A1-A2 in Appendix B are satisfied, Assumption A3 holds, $\lambda/\big[m^*T \delta_T \big]\rightarrow\infty$, $\lambda/\big[J^2_{\emph{min}}I_{\emph{min}}\big]\rightarrow 0$ \;as\; $T\rightarrow\infty$.
Then, we have 
\begin{enumerate}
\item[] $P(\widehat{m}=m^*)\rightarrow 1$ \;\;as\;\; $T\rightarrow\infty$;
\item[]
$P\Big(\max_{1\le j\le m^*}\big|\widehat{T}_j-T^*_j\big|\le T\delta_T\Big)\rightarrow 1$ \;\;as\;\; $T\rightarrow\infty$;
\item[] $\widehat{\balpha}_j-\balpha^*_j=O_p\Big([I_j^*]^{-1/2}\Big)$ \;\;for each\;\; $j=1,...,m^*+1$.
\end{enumerate}
\end{theorem}
\noindent\textit{Proof:} See Appendix C. 
\begin{remark}{
\rm
Our assumptions are weaker than the corresponding assumptions imposed by \cite{qian_su_2016}. First, 
we do not impose the two bounds involving $m^*$ that \cite{qian_su_2016} do in their Assumption 3. In particular, we allow $m^*$ to grow faster than $\log T$ as $T\rightarrow\infty$. Second, we do not impose the bound $\widehat{m}\le m_{\emph{max}}$, where $m_{\emph{max}}\le C\log T$, as \cite{qian_su_2016} do on page 1386  -- they use this bound in the proof of their Theorem 3.4 on recovering the correct number of breaks (see the statement and proof of their Lemma E.1 on page 1424). Thus, we do not restrict the range of~$\widehat{m}$ in our optimization problem to take advantage of the upper bound $m^*\le C\log T$; such a bound would typically be unknown in practice.
A similar phenomenon, where weaker assumptions are needed for the $\ell_0$-based estimators than for the
$\ell_1$-based estimators, has been observed in the regression setting; for example, this
phenomenon is discussed in the papers by \cite{bertsimas2016}, \cite{hazimeh2023grouped} and \cite{mazumder2023subset}.
}
\end{remark}
\begin{remark}{\rm
In their Section~3.2, \cite{qian_su_2016} impose conditions on tuning parameter~$\rho_T$, which controls the penalty on the total number of breaks in the information criterion that they use to determine the final estimator. We also impose conditions on~$\lambda$, which controls our penalty on the number of breaks. As the two estimators use these penalties differently, we cannot directly compare the conditions on~$\rho_T$ and $\lambda$. However, both sets of conditions are standard -- they are used to ensure that the penalty is neither too large nor too small, so that the correct number of breaks can be recovered with high probability. We note that the $\lambda$-range $m^*T \delta_T \ll \lambda \ll J^2_{\emph{min}}I_{\emph{min}}$, considered in Theorem~\ref{asympt.theorem}, is non-empty. Moreover, the width of this range tends to infinity as $T\rightarrow\infty$, because $J^2_{\emph{min}}I_{\emph{min}}\rightarrow\infty$ and $J^2_{\emph{min}}I_{\emph{min}}/[m^*T\delta_T]\rightarrow\infty$  under the conditions imposed in Assumption A3(ii).
}
\end{remark}

In the two remarks above, we compare our estimator to the following two-stage procedure of \cite{qian_su_2016}. First, a base GFL estimator is obtained for a range of values of the tuning parameter $\rho_T$; 
second, the final estimator is determined by selecting the tuning parameter using an information criterion that penalizes the number of breaks. In contrast, our approach does everything in one go, and avoids the estimation bias that comes from LASSO penalty. 

We can see that, in comparison to the base GFL estimator of \cite{qian_su_2016} 
our estimator enjoys better asymptotic properties. In particular, while \cite{qian_su_2016} show that the GFL estimator has at least as many breaks as the true model (see their Theorem 3.3), they do not establish a complimentary upper bound result. In contrast, we show that our estimator recovers the correct number of breaks with probability tending to one.

Next, we establish the asymptotic normality of our estimated regression coefficients. To state a clean result, we assume that $m^*$ is \emph{fixed} and nonzero. However, we note that this result can be extended to the general case as in \cite{qian_su_2016}, by imposing additional assumptions on~$m^*$ and stating the central limit theorem for pre-specified fixed-dimensional sub-vectors of coefficients.

We impose the following additional conditions, which are also required by \cite{qian_su_2016} in the analogous result for their estimator.

\noindent\textbf{Assumption A4}. 
\begin{enumerate}
\item[(i)] $\delta_T^{-1} I^{-1}_{\textrm{min}} \big[I^{1/2}_{\textrm{min}}T^{-1/2}(\log I_{\textrm{min}} )^{c_{\delta}/2}+1  \big]=O(1)$;          
\item[(ii)] $T\delta_T/I_{\textrm{min}}^{1/2}\rightarrow0$ as $T\rightarrow\infty$.
\end{enumerate}

Because our estimator recovers the correct number of breakpoints with probability tending to one, we follow the approach of \cite{qian_su_2016} and establish asymptotic normality for the estimator that solves optimization problem (2) with the restriction that the total number of breakpoints is exactly~$m^*$, i.e., $\sum_{t=1}^T z_t=m^*$. We write~$\widehat{\balpha}_{m^*}$ for the corresponding vector of estimated regression coefficients and observe that $\widehat{\balpha}_{m^*}=\big(\widehat{\mathbb{X}}^{\top}\widehat{\mathbb{X}}\big)^{-1}\widehat{\mathbb{X}}^{\top}Y$, where $\widehat{\mathbb{X}}=\text{diag}\big( (x_1,...,x_{\widehat{T}_1-1})^{\top},...,(x_{\widehat{T}_{m^*}},...,x_{T})^{\top} \big)$ and $Y=(y_1,...,y_T)^\top$. We also let $\Psi=\text{plim}D^{-1}\mathbb{X}^{\top}\mathbb{X}D^{-1}$ and $\Phi=\text{plim}D^{-1}\mathbb{X}^{\top}UU^{\top}\mathbb{X}D^{-1}$, where $D=\text{diag}\big({I^*_1}^{1/2}\mathbb{I}_p,...,{I^*_{m^*+1}}^{1/2}\mathbb{I}_p \big)$ with~$\mathbb{I}_p$ denoting the $p\times p$ identity matrix, $U=(u_1,...,u_T)^\top$, and $\mathbb{X}$ is defined analogously to $\widehat{\mathbb{X}}$ but using the true rather than the estimated breakpoints. 

\begin{theorem}
\label{CLT}
Let $\widehat{D}=\emph{diag}\big([\widehat{T}_1-\widehat{T}_0]^{1/2}\mathbb{I}_p,...,[\widehat{T}_{T}-\widehat{T}_{m^*}]^{1/2}\mathbb{I}_p \big)$ and suppose that Assumptions A1-A4 hold. Then, $\widehat{D}\big(\widehat{\balpha}_{m^*}-\balpha^*\big)\xrightarrow[]{d}N\big(\mathbf{0},\Psi^{-1}\Phi\Psi^{-1}\big)$.
\end{theorem}
\noindent\textit{Proof:} See Appendix D.

We note that Theorem~\ref{CLT} is a direct consequence of Theorem 3.6 in \cite{qian_su_2016} on the asymptotics of their post-LASSO estimator of regression coefficients. We also note that the asymptotic variance matrix can be estimated by replacing $\Psi$ and $\Phi$ with estimates obtained using $\widehat{D},\widehat{\mathbb{X}}$ and $\widehat{U}=Y-\widehat{\mathbb{X}}\widehat{\balpha}_{m^*}$.

\section{Monte-Carlo simulations}\label{sec:sims}

In this section, we investigate the finite sample properties of our proposed approach, focusing on the number of identified breaks.

In order to compare the new estimator to the GFL approach, we follow \citet{qian_su_2016} and use the same data generating process as in \eqref{eq:regression} with the following cases of interest: the case of no breaks, the case of one break, and the case of many breaks. We compare MIO and GFL methods with classical approaches used by  \citet{bai2003computation}, namely, BIC and LWZ information criteria \cite[see][]{liu/etal:97}, and the sequential method  SEQ of \citet{bai1998estimating}. 

BIC and LWZ are natural comparisons as they choose the number of breaks using information criteria. Specifically, they minimize $BIC(m)=\ln{(SSR(\widehat{T}_1,\dots,\widehat{T}_m)/T)}+p^{\star}\ln(T)/T$ and $LWZ(m)=\ln{(SSR(\widehat{T}_1,\dots,\widehat{T}_m)/(T-p^{\star}))}+(p^{\star}/T)0.299(\ln(T))^{2.1}$ over $m$, where $p^{\star}=(m+1)p+m$ and $SSR(\widehat{T}_1,\dots,\widehat{T}_m)$ is the sum of squared residuals from estimating \eqref{eq:regression} and  the break dates $\widehat{T}_1,\dots,\widehat{T}_m$ are obtained by minimizing $SSR(T_1,\dots,T_m)$ over all possible combinations of $T_1,\dots,T_m$. 
The commonly used sequential method  SEQ is  based on first testing for the presence of breaks using the UDmax test of \cite{bai1998estimating}, and then sequentially testing the null hypothesis of $l$ breaks against the alternative of $l+1$ breaks from $l=1$. 

We generate 500 replications for each problem instance described below. For each problem instance, we solve optimization problem~(\ref{l0-penal}) 
in the MIO formulation (\ref{est.eq1})-(\ref{est.eq4}). We use a widely available and highly powerful Gurobi 9.5 Solver on one core of AWS EC2 r5.16xlarge (64 cores, 512 GB RAM), with the time limit for each problem set to 450 seconds. 

We solve the MIO problem $N$ times for different values of $\lambda \in \{\lambda_1, \lambda_2, ..., \lambda_N\}$ and find the final estimator via the information criterion, as in \cite{qian_su_2016}. More specifically, we solve  $\min_{\lambda_j \in \{\lambda_1, \lambda_2, ..., \lambda_N\}} \log\frac{1}{T}\sum_{t=1}^{T}(y_{t} - \widehat{\bbeta}_{\lambda_j,t}^{\top}x_t)^2 + \frac{1}{\sqrt{T}} p (\hat{m}_j+1)$, where $\widehat{\bbeta}_{\lambda_j,t}$ is the solution corresponding to $\lambda_j$ and $\hat{m}_j$ is the number of breaks in this solution.\footnote{Python codes and data for simulations and applications are available on the corresponding author's web page and Github.} In practice, we ensure that our grid $\{\lambda_1, \lambda_2, ..., \lambda_N\}$ is sufficiently dense to account for {every} possible number of estimated breakpoints between zero and a
computationally reasonable upper bound~$B$; for example, $B=25$. 
We recommend the following approach: (a) if the information criterion is minimized at a~$\lambda$ for which $\hat{m}_{\lambda}<B$, then we stop and use $\widehat{\bbeta}_{\lambda,t}$ as our final estimator; (b) if the information criterion is minimized at a~$\lambda$ with $\hat{m}_{\lambda}=B$, then we increase $B$ to $1.2B$ and re-evaluate the criterion. We continue this process until the minimum of the criterion corresponds to~$\hat{m}_{\lambda}<B$.

We report results for the case of one break and many breaks. The case with zero breaks, which confirms our method is competitive in terms of correct detections, is reported in Appendix E. Computational details including boxplots of optimality gaps and run-times are available in an online Supplement.


\subsection{The case of one break}

In this section we use the following DGP variations:
\begin{equation}
    y_t=\beta_tx_t+u_t,
\end{equation}
where
\begin{enumerate}
    \item $\beta_t = \textbf{1}\{T/2 < t \leq T\}$, $x_t \sim i.i.d. N(0,1)$, $u_t \sim i.i.d. N(0, \sigma^2_u)$
    \item $\beta_t = \textbf{1}\{T/2 < t \leq T\}$, $x_t \sim i.i.d. N(0,1)$, $u_t = \sigma_u v_t$ with $v_t = 0.5v_{t-1}+\epsilon_t$, $\varepsilon_t \sim N(0,0.75)$
    \item $\beta_t = \textbf{1}\{T/2 < t \leq T\}$, $x_t=0.5x_{t-1}+\eta_t$, $\eta_t\sim i.i.d. N(0,0.75)$, $u_t \sim i.i.d. N(0,\sigma_u^2)$
    \item  $\beta_t = \textbf{1}\{T/2 < t \leq T\}$, $x_t=0.5x_{t-1}+\eta_t$, $\eta_t\sim i.i.d. N(0,0.75)$, $u_t = \sigma_u \sqrt{h_t}\varepsilon_t$, $h_t = 0.05 + 0.05 u_{t-1}^2 + 0.9h_{t-1}$, $\varepsilon_t \sim i.i.d. N(0,1)$
    \item $\beta_t = \textbf{1}\{T/2 < t \leq T\}$, $x_t=0.5x_{t-1}+\eta_t$, $\eta_t\sim i.i.d. N(0,0.75)$, $u_t = \sigma_u v_t$ with $v_t = \varepsilon_t + 0.5 \varepsilon_{t-1}$, $\varepsilon_t \sim i.i.d. N(0,0.8)$ 
    \item $\beta_t = 0.2\textbf{1}\{1 < t \leq T/2\} + 0.8\textbf{1}\{T/2 < t \leq T\}$, $x_t = y_{t-1}$, $u_t \sim i.i.d. N(0, \sigma_u^2)$,
\end{enumerate}
and the true parameter values for $\sigma_u$ are the same as in the last subsection. 

Table \ref{tab1_1break} reports the percentage of replications that detected the correct number of breaks (one) and the accuracy of the break date detection for MIO, GFL, BIC, LWZ and SEQ. 
Column \emph{pce} contains the percentage of correct detections, 
column \emph{hd/T} reports the Hausdorff distances, divided by $T$,  between estimated break date and true break date (conditional on the correct estimation of the number of breaks) and serves as a measure of the overall accuracy of break date estimation.

It can be seen from Table \ref{tab1_1break} that MIO and GFL produce comparable results. MIO tends to outperform GFL in terms of $pce$ in larger samples. In fact, when $n=500$, MIO outperforms all the competitors in terms of $pce$, except for DGP-4 in the low signal-to-noise regime ($\sigma_u=1.5$), in which all estimators behave extremely poorly regardless of the sample size.  
Additional results, not reported for brevity, suggest that MIO tends to overestimate the number of breaks in small samples while GFL tends to underestimate the number of breaks in small samples. 

In terms of accuracy of the break date detection, MIO outperforms GFL in all sample sizes but only in high signal-to-noise environments ($\sigma_u=0.5$). In the mid to low signal-to-noise regimes the performance of the five estimators is mixed but MIO shows a superior performance in larger samples. 

Overall, the percentage of correct detections decreases for all methods when the noise variance, $\sigma_u$, increases. Both correct detection and accuracy improve as the sample size $T$ increases.

\begin{table}[!htbp]
\centering
\footnotesize
\caption{Correct detections and estimation accuracy: one break.}\label{tab1_1break}
\begin{tabular}{ccccccccccccc} \toprule
 & & &MIO& &GFL& &BIC& &LWZ& &SEQ& \\\hline
 &$\sigma_u$&$T$&pce&hd/T&pce&hd/T&pce&hd/T&pce&hd/T&pce&hd/T\\\hline
DGP-1&0.5&100&94.2&1.2&98.6&1.7&96.4&1.3&100&1.3&88.6&1.2\\
 &&200&99.4&0.6&99.2&0.8&98.8&0.6&100&0.6&95.2&0.6\\
 &&500&100&0.2&99.8&0.3&99.4&0.2&100&0.2&96.8&0.2\\
 &1&100&92&4.4&95&3.6&96&4.1&80&3.8&90.2&4.2\\
 &&200&99&1.9&99.4&1.9&98.2&1.9&98.8&1.9&96&1.9\\
 &&500&100&0.8&99.2&0.8&99.4&0.8&100&0.8&96.6&0.7\\
 &1.5&100&63.6&7.6&64.8&5.3&70.6&6.7&28&5.7&71.4&7.0\\
 &&200&86.2&3.7&86.2&3.1&94&3.8&55.6&3.3&93.8&3.9\\
 & &500&99.6&1.6&98.6&1.4&99.4&1.6&97.8&1.6&97&1.6\\\hline
DGP-2&0.5&100&91.2&1.0&98.4&1.5&97.8&1.0&99.8&1.0&88.8&1.1\\
 &&200&97&0.6&98.6&0.8&97.2&0.6&100&0.6&93.4&0.6\\
 &&500&99.8&0.2&99.4&0.3&98.8&0.2&100&0.2&95.6&0.2\\
 &1&100&88.2&3.7&95.2&3.0&96&3.6&81.2&3.2&88.6&3.5\\
 &&200&96.2&1.7&99.2&1.6&97.2&1.7&98.8&1.6&93.2&1.6\\
 &&500&99.6&0.7&99.4&0.7&98.6&0.7&100&0.7&95.4&0.7\\
 &1.5&100&60.8&7.0&63.4&4.9&69.8&6.4&27.4&4.8&72.6&6.4\\
 &&200&83.4&3.5&84&3.0&93.2&3.5&55.8&2.8&91.8&3.7\\
 & &500&99.4&1.5&98.8&1.4&99.2&1.6&96.4&1.5&96&1.6\\\hline
DGP-3&0.5&100&93.8&1.2&98.8&1.6&96.6&1.2&100&1.2&88.4&1.2\\
 &&200&99.2&0.6&100&0.8&98.4&0.6&100&0.6&94.8&0.6\\
 &&500&100&0.2&99.8&0.3&99&0.2&100&0.2&94.4&0.2\\
 &1&100&89.2&4.5&93.4&3.7&94&4.1&77.2&3.9&86.4&4.2\\
 &&200&99.4&2.0&99.4&1.8&97.4&2.0&98.6&2.0&94.4&1.9\\
 &&500&100&0.7&99.8&0.8&99&0.7&100&0.7&94.2&0.7\\
 &1.5&100&60.6&8.2&61.2&5.6&67&7.4&22.4&5.8&69.8&8.0\\
 &&200&87.8&4.3&87.2&3.1&93.4&4.4&58&3.7&93.8&4.4\\
 & &500&99.6&1.7&99&1.4&99.2&1.7&98.8&1.7&95.6&1.7\\\hline

DGP-4&0.5&100&95.4&0.8&99.6&1.4&97.2&0.8&99.6&0.8&90.4&0.8\\
 &&200&99.6&0.5&99.6&0.7&97.6&0.5&100&0.5&93.4&0.5\\
 &&500&100&0.2&99.8&0.3&99.6&0.2&100&0.2&96.6&0.2\\
 &1&100&90&3.8&95.4&3.3&93.8&3.7&84.4&3.5&87.6&4.0\\
 &&200&97.6&1.9&99.4&1.8&97.2&2.0&99&1.9&94&2.0\\
 &&500&99.8&0.8&98.8&0.8&99.6&0.8&100&0.8&96.6&0.8\\
 &1.5&100&23.6&21.7&19.8&12.6&24&13.7&4.4&14.5&31.8&12.9\\
 &&200&15.4&16.8&15&14.1&23.6&11.8&1.2&8.0&31.8&10.0\\
 & &500&3.6&29.3&4.4&27.5&16.2&15.8&0.4&32.6&21.4&10.6\\\hline
DGP-5&0.5&100&95.6&1.1&97.8&1.5&97&1.1&100&1.2&86.4&1.1\\
 &&200&98.6&0.6&99.4&0.9&98.6&0.6&100&0.6&92.8&0.6\\
 &&500&100&0.2&99.2&0.3&99.2&0.2&100&0.2&96.4&0.2\\
 &1&100&93.6&4.1&96.2&3.7&95.8&4.2&80.6&4.1&87.8&4.4\\
 &&200&98.4&1.9&99.6&1.9&98.8&1.8&99.6&1.9&92.4&1.8\\
 &&500&100&0.7&99.2&0.7&99&0.7&100&0.7&96&0.7\\
 &1.5&100&63.4&7.9&64.2&5.9&70.2&7.4&27.6&6.6&72.2&8.1\\
 &&200&87&4.4&87.4&3.4&95&4.1&58.4&3.6&92&4.1\\
 & &500&99.6&1.5&99&1.4&98.8&1.5&98.8&1.5&96.2&1.4\\\hline
DGP-6&0.5&100&65&8.1&64.6&8.4&69.8&7.0&28.8&5.9&70&7.6\\
 &&200&93&4.2&93&5.8&97&4.2&71.2&4.1&91.8&4.2\\
 &&500&100&1.5&97.6&2.5&99.2&1.5&99.6&1.5&93.4&1.5\\
 &1&100&65&8.1&64.6&8.4&69.8&7.0&28.8&5.9&70&7.6\\
 &&200&93&4.2&93&5.8&97&4.2&71.2&4.1&91.8&4.2\\
 &&500&100&1.5&97.6&2.5&99.2&1.5&99.6&1.5&93.4&1.5\\
 &1.5&100&65&8.1&64.6&8.4&69.8&7.0&28.8&5.9&70&7.6\\
 &&200&93&4.2&93&5.8&97&4.2&71.2&4.1&91.8&4.2\\
 & &500&100&1.5&97.6&2.5&99.2&1.5&99.6&1.5&93.4&1.5\\\hline
   \bottomrule
\end{tabular}
\end{table}

\subsection{The case of many breaks}

We follow the simulation design of \citet{qian_su_2016} and consider
\begin{equation}
    y_t=\beta_tx_t+u_t,
\end{equation}
where $x_t \sim i.i.d. N(0,1)$, $u_t \sim i.i.d. N(0, \sigma^2_u)$,
\begin{equation*}
    \beta_t=\begin{cases}
        0, & \Delta(2i)+1\leq t<\Delta(2i+1) \\
        1, & \Delta(2i+1)+1\leq t<\Delta(2i+2) \\
    \end{cases}, \ i=0,1,\dots,R/2.
\end{equation*}
For the first design (DGPn-1),  we fix the length of the regime $\Delta=30$ and allow a different number of regimes $R\in\{6,10,20\}$. For the second design (DGPn-2),  we fix the number of regimes $R=10$ and allow different regime lengths by varying $T\in\{150,300,600\}$.

Table \ref{tab_many_breaks} reports the percentage of correct detections $pce$ and scaled Hausdorff distance $hd/T$ for each  number of breaks $R$ and each sample size $T$. We report $pce$ only for MIO and GFL because the classical methods of \citet{bai1998estimating,bai2003computation} do not permit 9 and 19 breaks for the selected sample sizes. We can see from the table that MIO uniformly dominates GFL in all cases in terms of $pce$, and the better performance is particularly noticeable in cases of higher error variance ($\sigma_u=0.5$).\footnote{Our results for GFL are different from (and somewhat worse than) those reported by \cite{qian_su_2016}. We used the \emph{regsc} package provided by Junhui Qian on his webpage and the same options for all simulations with lambda=NULL and method=``ic'' as the defaults. We note that our MIO results dominate both our version of GFL and the original version.} Also, the accuracy of MIO is higher for most cases in terms of the Hausdorff distance.

\begin{table}[!h]
\centering
\footnotesize
\caption{Correct detections and estimation accuracy: many breaks.}\label{tab_many_breaks}
\begin{tabularx}{0.6\textwidth}{cccccccc} \toprule
&&&&MIO&&GFL&\\\hline
DGPn-1&$\sigma_u$&R&T&pce&hd/T&pce&hd/T\\\hline
 &0.2&6&180&98.8&0.6&86.6&0.6\\
 &&10&300&98.6&0.5&76.4&0.5\\
 &&20&600&100.0&0.4&56.4&0.3\\
 &0.5&6&180&99.2&1.9&37.8&2.0\\
 &&10&300&94.8&1.4&26.8&3.5\\
 & &20&600&27.0&1.0&1.8&1.6\\\hline
 & & & &MIP& &GFL& \\\hline
DGPn-2&$\sigma_u$&R&T&pce&hd/T&pce&hd/T\\\hline
 &0.2&10&150&95.8&1.1&66.6&1.0\\
 &&10&300&99.2&0.5&78.0&0.5\\
 &&10&600&100.0&0.2&82.8&0.3\\
 &0.5&10&150&43.2&2.8&12.4&3.8\\
 &&10&300&94.4&1.5&19.6&4.4\\
 & &10&600&100.0&0.8&23.0&2.1\\\hline
   \bottomrule
\end{tabularx}
\end{table}

\section{Real data examples}\label{sec:empirics}

\subsection{Level shifts in US real interest rate}

In this subsection, we consider the U.S. real interest rate time series from 1961Q1 to 1986Q3 used by \cite{garcia1996analysis} and \cite{bai2003computation}. Of interest is a simple level shift model which can be written as follows
\begin{equation}
    y_t=\mu_j+u_t, \ j=1,\dots,m^*+1,
\end{equation}
where $y_t$ is the real interest rate obtained using the U.S. 90-day Treasury bill rate and a quarterly inflation rate series constructed from the U.S. CPI.

\begin{figure}[h]%
\begin{center}%
\includegraphics[width=0.6\linewidth]{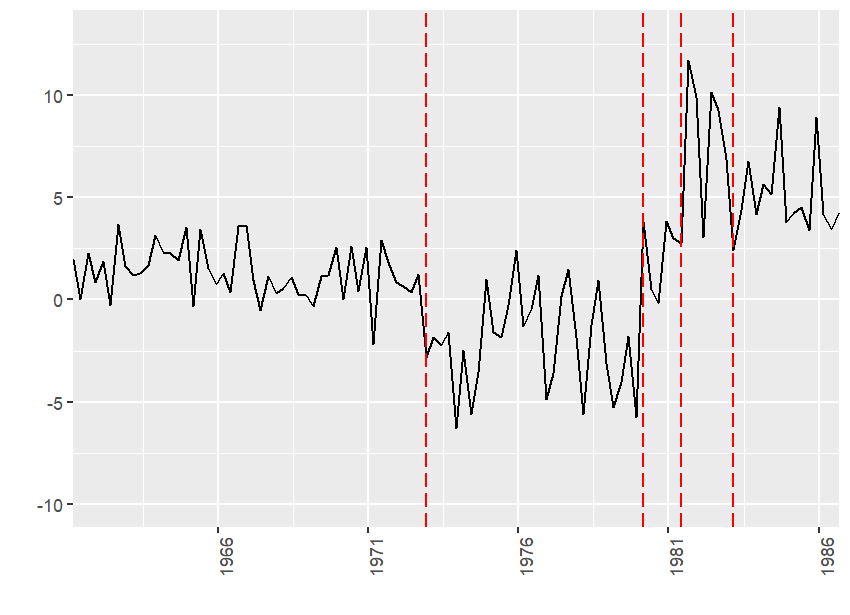}
\end{center}%
\caption{US real interest rate; 1961Q1 -- 1986Q3.}
\label{fig_rate}
\end{figure}

Table \ref{tab_emp_1} reports the estimation results using the five methods. The breakpoints detected by MIP are depicted in Figure \ref{fig_rate}. The MIP and GFL methods detect 4 breaks while the classical methods BIC, LWZ and SEQ detect 2, 0 and 3 breaks, respectively. The most common break date is 1972Q4 which can be tied to the oil crisis, while the breaks in the 1980s can be associated with Paul Volcker's attempt to lower the inflation in the US and with the start of what became known as the Great Moderation, that is, a decrease in the variance of the error term. 
For example, \citet{walsh1987testing} noted that in early 1980, real interest rates may have reacted to a change in the Federal Reserve operating procedures implemented in late 1979.

We can see that the level breaks detected by MIP and GFL in the 1980s are very close to one another. This questions the assumptions of BIC, LWZ and SEQ about the minimal length of the regime, which may affect these methods' performance.

\begin{table}[!h]
\centering
\scriptsize
\caption{Estimated break dates, US real interest rate}\label{tab_emp_1}
\begin{tabularx}{0.9\textwidth}{cccccccc} \toprule
&$\hat{m}$&Dates&&&&&\\\hline
MIP&4&&1972Q4&1980Q1&&1981Q3&1983Q1\\
GFL&4&&1972Q4&1980Q1&1980Q4&1981Q3&\\
BIC&2&&1972Q4&&1980Q4&&\\
LWZ&0&&&&&&\\
SEQ (trim=0.1)&3&1967Q1&1972Q4&&1980Q4&&\\
\hline
   \bottomrule
\end{tabularx}
\end{table}

Our finding of the additional break in 1983Q1 provides support to \cite{bauwens/etal:15} who argue that a distinct regime starting in 1983Q1 is characterized by the lowest variability, the highest persistence and an average growth rate between the other regimes, and thus best defines the start of the Great Moderation. This finding also contributes to the literature on what caused the high real interest rates in the U.S.~in the early 1980s. Specifically, it puts more weight on the Federal Reserve's attempts in late 1982 to de-emphasize monetary aggregates, described by \cite{walsh1987testing}. As argued by \cite{huizinga1986monetary},  this suggests that it is the monetary policy, not fiscal deficits, that bears the responsibility for high U.S.~real interest rates in the early 1980s.

\subsection{Inventory adjustment model}

In this subsection, we consider a business statistics application from inventory modeling. A classic model in this literature is 
\[z_t  = \mu +(1-\alpha) z_{t-1} + \epsilon_t, \]
where $\epsilon_t$ is white noise, $z_t$ is a disequilibrium variable, e.g., inventory-to-shipment ratio, and $\alpha\in[0,1]$ is the speed of adjustment \cite[see, e.g., ][]{caballero/engel:93, jorda1999random, jorda/marcellino:04}. A problem  with explaining the inventory dynamic is that the empirical values of the adjustment speed are much lower than predicted by economic theory. For example, estimates of $\alpha$ referenced by \cite{jorda/marcellino:04} and \cite{jorda1999random} are between 5\% and 29\% for monthly data suggesting that it takes over a year to close 95\% of a given disequilibrium.

One solution proposed by \cite{jorda1999random} is to recognise that data aggregation  happens at irregular intervals -- different from the reporting frequency -- and affects $\alpha$. For example, \cite{jorda/marcellino:04} estimate a Markov switching model with three regimes representing what they call zero, one and two original-time adjustments, that is, three types of time-varying aggregation frequency. The values of $1-\alpha$ they find are 0.99, 0.72 and 0.58. Because of the limitation on the minimum number of observations in each regime and the small sample,  their number of regimes is fixed at three. 

\begin{figure}[h]%
\begin{center}%
\includegraphics[width=0.6\linewidth]{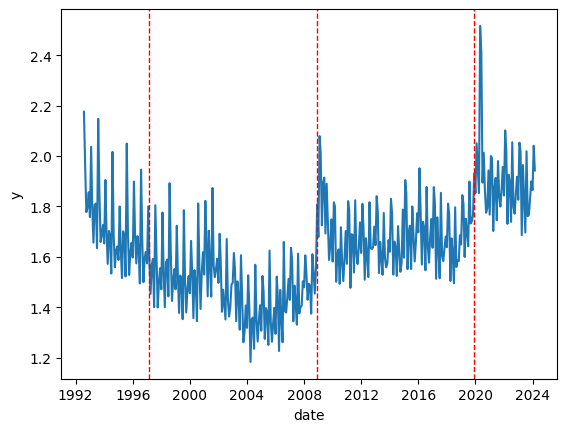}
\end{center}%
\caption{Inventory-to-shipment ratio; 1992 Jan -- 2023 May.}
\label{fig:ratio}
\end{figure}

We re-estimate the model without that restriction. We use the data on total manufacturing from the Census Bureau’s monthly survey ``Manufacturers’ Shipments, Inventories and Orders''. This is the same data source as used  by \cite{jorda/marcellino:04} and \cite{jorda1999random} but a larger product category (they used glass containers which are no longer available separately), nominal volumes and a larger sample ranging from January 1992 to May 2023. Figure \ref{fig:ratio} shows the inventory-to-shipment ratio for the entire sample of 380 observations. Similar to \citet[][Figure 1]{jorda/marcellino:04}, the data is not seasonally adjusted, which is preferable when investigating inventory dynamics. 

When we apply our method to this data, we find four regimes, which are delineated in Figure \ref{fig:ratio} by the dashed lines and correspond to January 1997, October 2008 and November 2019. The estimated values of $(1-\alpha)$ are 0.142, 0.135, 0.094, and 0.041, suggesting 
the adjustment speed $\alpha$ of 85.8-95.9\%, which is much faster than in previous studies.  
A simple calculation used by \citet[][p.~392]{jorda1999random} suggests that 95\% of a disequilibrium is replenished within $T=30 \log(0.05)/\log(0.142) =46$ days, i.e., within a month and a half -- a more realistic estimate.  

For comparison, the GFL estimator produced zero breaks; BIC and SEQ resulted in four breaks but the estimates of $(1-\alpha)$ include negative values, which are infeasible and, thus, not reported. 
LWZ produced three  breaks in July 1997, October 2008 and May 2019, and the estimates of $(1-\alpha)$ are 0.195, 0.140, 0.133 and 0.036. Three of these estimates are very close to ours while the highest value of 0.195 means that the longest period required to close 95\% of a disequilibrium is about two months, i.e., only two weeks longer than our estimate.

\citet{jorda/marcellino:04} hypothesize that these regimes reflect time aggregation, which in effect scales the original sampling frequency. 
If this is the case,  one can make an argument similar to \citet[][pp.~884-885]{jorda/marcellino:04} that the estimates of $\alpha$ corresponding to the different aggregation scales should relate to each other. 
A smaller estimate of $1-\alpha$ is expected to be close to a power of a larger estimate. 
It turns out that our largest and smallest estimated values of $1-\alpha$ (i.e., 0.142 and 0.041) agree with this prediction: $0.142^2 = 0.02$, which is not far from 0.041, or, equivalently, the implied $\alpha$ derived from $(1-\alpha)^2=0.041$ is 0.8, which is not far from 0.858. We note that a similar relationship exists in the LWZ estimates: the implied $\alpha$ derived
from $(1 -\alpha)^2 = 0.036$ is 0.81, which coincides with $1-0.195=0.805$. The intermediate values of $1-\alpha$ found by our method and by LWZ do not seem to obey this relationship, suggesting that these regimes may have a different nature. Nonetheless, the above findings can serve as evidence in favor of the hypothesis of \cite{jorda/marcellino:04} for much higher speeds of adjustment than previously thought. 

\section{Conclusion}\label{sec:conclude}

We propose a new way of handling change-points in econometrics based on computational advances in mixed integer optimization and we work out statistical properties for the estimator of the number of breaks, break locations and the regression coefficient in one step. The approach shows remarkable adaptivity and versatility in that it has similar or better asymptotic properties than the LASSO-based alternatives under somewhat weaker assumptions. In simulations, the proposed method shows improved performance especially under many breaks scenarios. In empirical applications, it provides additional insights, offering a new and robust way to obtain evidence on the number and location of breaks, and the corresponding regime characteristics without the restrictive assumptions on the regime duration. We demonstrate this for two classic examples in economics.

\bibliographystyle{apalike}
\bibliography{references}

\section*{Appendices}
\appendix
\subsection*{Appendix A: MIO formulation (\ref{est.eq1})-(\ref{est.eq4})}\label{app:A}

In this section, we show that optimization problems (\ref{l0-penal}) and (\ref{est.eq1})-(\ref{est.eq2}), (\ref{est.eq4}) are equivalent as long as the Big-M constant $M$ is sufficiently large. 

We recall our \(\ell_0\)-penalized regression problem:
\begin{equation}
\label{l0-penal2}
\widehat{\boldsymbol{\beta}}
=\arg\min_{\boldsymbol{\beta}} 
\sum_{t=1}^{T}
\bigl(y_t - \boldsymbol{\beta}_t^{\top} x_t\bigr)^2 
+ \lambda\sum_{t=2}^{T}\mathbf{1}\{\boldsymbol{\beta}_t \neq \boldsymbol{\beta}_{t-1}\}
\end{equation}
Suppose that $M\ge \max_{0\le t\le T-1}\|\widehat{\bbeta}_{t+1}-\widehat{\bbeta}_{t}\|_{\infty}$, and consider the following MIO formulation:
\begin{subequations}
\begin{align} 
\min_{\boldsymbol{\beta}_t,\,z_t} \quad 
& \sum_{t=1}^{T}(y_t - \boldsymbol{\beta}_t^{\top} x_t)^2 
+ \lambda\sum_{t=1}^{T-1}z_t, 
\label{est:eq1}\\
\text{s.t.} \quad 
& -M\mathbf{e}\,z_{t} \leq \boldsymbol{\beta}_{t+1} - \boldsymbol{\beta}_{t} 
\leq M\mathbf{e}\,z_{t}, \quad t=1,\dots,T-1,
\label{est:eq2}\\
& z_{t} \in \{0,1\}, \quad t=1,\dots,T-1.
\label{est:eq4}
\end{align}
\end{subequations}

When \(z_t = 0\), constraint~(\ref{est:eq2}) becomes
\(
   -M\mathbf{e}\cdot 0 \;\le\; \boldsymbol{\beta}_{t+1} - \boldsymbol{\beta}_{t} 
   \;\le\; M\mathbf{e}\cdot 0,
\)
and hence \(\boldsymbol{\beta}_{t+1} = \boldsymbol{\beta}_{t}\).
When \(z_t = 1\), the constraint becomes
\(
   -M\mathbf{e} \;\le\; \boldsymbol{\beta}_{t+1} - \boldsymbol{\beta}_{t} 
   \;\le\; M\mathbf{e},
\)
or $\|\boldsymbol{\beta}_{t+1} - \boldsymbol{\beta}_{t}\|_{\infty}\le M$, allowing 
\(\boldsymbol{\beta}_{t+1}\) to differ from \(\boldsymbol{\beta}_{t}\). 
Conversely, \(\boldsymbol{\beta}_{t+1} = \boldsymbol{\beta}_{t}\) implies $z_t=0$, because otherwise $z_t$ would make a positive contribution to the objective function in~(\ref{est:eq1}).  Similarly,  \(\boldsymbol{\beta}_{t+1} \ne \boldsymbol{\beta}_{t}\) implies $z_t=1$ due to the constraints in~(\ref{est:eq2}). 

Consequently, the term \(\sum_{t=1}^{T-1} z_{t}\) in the objective function of the MIO formulation maps directly to the term
\(\sum_{t=2}^{T} \mathbf{1}\{\boldsymbol{\beta}_{t} \neq \boldsymbol{\beta}_{t-1}\}\) 
in the original problem~(\ref{l0-penal2}), under the additional constraint that $\max_{0\le t\le T-1}\|{\bbeta}_{t+1}-{\bbeta}_{t}\|_{\infty}\le M$.


%
%

\subsection*{Appendix B: Theoretical assumptions}

We now state Assumptions A1 and A2 from \cite{qian_su_2016}. We denote by~$\mu_{\text{max}}$ and~$\mu_{\text{min}}$ the largest and the smallest eigenvalues, respectively, of a symmetric matrix~$A$.

\noindent\textbf{Assumption A1}.
\begin{enumerate}
\item[(i)] $\{(x_t,u_t)\}$ is a strong mixing process with mixing coefficients $\alpha(\cdot)$ satisfying $\alpha(\tau)\le c_{\alpha}\rho^{\tau}$ for some $c_{\alpha}>0$ and $\rho\in(0,1)$. $E(x_tu_t)=0$ for each~$t$.
\item[(ii)] Either one of the following two conditions is satisfied: (a) $\sup_{t\ge1}E\|x_t\|^{4q}<\infty$ and $\sup_{t\ge1}E|u_t|^{4q}<\infty$ for some $q>1$;(b) There exist positive constants $c_{xx}$, $c_{xu}$ $c_{uu}$ such that $\sup_{t\ge 1} E\big[\exp(c_{xx}\|x_t\|^{2\gamma}) \big]<\infty$, $\sup_{t\ge 1} E\big[\exp(c_{xu}\|x_tu_t\|^{\gamma}) \big]<\infty$ and $\sup_{t\ge 1} E\big[\exp(c_{uu}|u_t|^{2\gamma}) \big]<\infty$ for some $\gamma\in(0,\infty]$. The case $\gamma=\infty$ is understood as uniform boundedness of $\|(x_t,u_t)\|$.
\end{enumerate}
\noindent\textbf{Assumption A2}.
\begin{enumerate}
\item[(i)] There exist two positive constants $\underline{c}_{xx}$ and $\overline{c}_{xx}$ and a positive sequence $\{\delta_T\}$ declining to zero as $T\rightarrow\infty$ such that
\begin{eqnarray*}
\underline{c}_{xx}&\le&\inf_{r-s\ge T\delta_T} \mu_{\textrm{min}}\Big(\frac1{r-s}\sum_{t=s}^{r-1} E(x_tx_t^{\top}) \Big) \\
&\le& \sup_{r-s\ge T\delta_T} \mu_{\textrm{max}}\Big(\frac1{r-s}\sum_{t=s}^{r-1} E(x_tx_t^{\top}) \Big) \le \overline{c}_{xx}.
\end{eqnarray*}
\item[(ii)] $T\delta_T$ satisfies one of the following two conditions: (a) $T\delta_T\ge c_vT^{1/q}$ for some $c_v>0$ is A1(ii.a) is satisfied; (b) $T\delta_T\ge c_v(\log T)^{(2+\gamma)/\gamma}$ for some $c_v>0$ if A1(ii.b) is satisfied.
\end{enumerate}
We note that inequality $\sup_{t\ge 1}E\big[\exp(c_{uu}|u_t|^{2\gamma}) \big]<\infty$ is omitted from assumption A1(ii.b) of \cite{qian_su_2016}; however, it is needed in their proofs. This inequality is the counterpart of inequality $\sup_{t\ge1}E|u_t|^{4q}<\infty$ in assumption A1(ii.a), and is required to control the tail behavior of the error term~$u_t$.

\subsection*{Appendix C: Proof of Theorem ~\ref{asympt.theorem}}

\textbf{Preliminaries}

First, we state some existing results that we will use in the proof of Theorem~\ref{asympt.theorem}.

\begin{lemma}
\label{lem.suppl}
    Suppose that Assumptions A1 and A2 hold. Then,
    \begin{enumerate}
        \item[\emph{(i)}] $\sup\limits_{r-s\ge T\delta_T}\mu_{\emph{max}}\big(\frac{1}{r-s}\sum_{t=s}^{r-1} x_tx_t^{\top}\big)\le \overline{c}_{xx} + o_p(1)$; 
        \item[\emph{(ii)}] $\inf\limits_{r-s\ge T\delta_T}\mu_{\emph{min}}\big(\frac{1}{r-s}\sum_{t=s}^{r-1} x_tx_t^{\top}\big)\ge \underline{c}_{xx}+o_p(1)$;
        \item[\emph{(iii)}] $\sup\limits_{r-s\ge T\delta_T}\big\|\frac{1}{\sqrt{r-s}}\sum_{t=s}^{r-1} x_tu_t\big\|=O_p\big([\log T]^{c_{\delta}/2}\big)$;
        \item[\emph{(iv)}] $\sup\limits_{0<r-s< T\delta_T}\sum_{t=s}^{r-1} u^2_t=O_p(T\delta_T)$.
    \end{enumerate}
\end{lemma}
Parts (i) and (ii) of Lemma~\ref{lem.suppl} are established in Lemma A.3 of \cite{qian_su_2016}; part (iii) is established in their Lemma A.4; part (iv) -- in the proof of their Lemma E1 (page 1425).

\textbf{Proof of Theorem~\ref{asympt.theorem}}
\begin{proof} 
Throughout the proof, we use $c_1$, $c_2$, $c_2$, ... to denote positive universal constants. Given a vector $\bbeta=(\bbeta_1^{\top},...,\bbeta_T^{\top})^{\top}$, with $\bbeta_j\in\mathbb{R}^p$, we define
$$
Q(\bbeta)=\sum_{t=1}^T\big(y_t-\bbeta_t^{\top}x_t\big)^2.
$$
Note that
\begin{equation}
\label{Q.eq.prf}
Q(\widehat{\bbeta})-Q({\bbeta}^*)=\sum_{t=1}^T\Big[(\widehat{\bbeta}_t-\bbeta^*_t)^{\top}(x_tx_t^{\top})(\widehat{\bbeta}_t-\bbeta^*_t)-
2(\widehat{\bbeta}_t-\bbeta^*_t)^{\top}x_tu_t\Big].
\end{equation}
We will prove the three claims of Theorem~\ref{asympt.theorem} in sequence.

\medskip

\noindent\textbf{Claim 1: }$\mathbf{P}(\widehat{m}=m^*)\rightarrow 1$. Using the combined set of the true and estimated breakpoints, $\{T^*_j\}\cup\{\widehat{T}_j\}$, we can divide the time interval index set into a collection of consecutive time intervals,  $\{1,...,T\}=\cup_{k} C_k$, so that on each such interval both the estimated and the true regression coefficients stay constant, i.e., neither~$\widehat{\bbeta}_t$ nor~$\bbeta^*_t$ change their values for $t\in C_k$, provided that the interval~$C_k$ is fixed. We let $\widehat{h}_k=\widehat{\bbeta}_t-\bbeta^*_t$ for $t\in C_k$, noting that this definition does not depend on the specific $t$ as long as $t\in C_k$. Thus, we can rewrite equation~(\ref{Q.eq.prf}) as follows:
\begin{equation}
\label{Q.eq.rwr}
Q(\widehat{\bbeta})-Q({\bbeta}^*)=\sum_k\left[\widehat{h}_k^{\top}\Big(\sum_{t\in C_k}x_tx_t^{\top}\Big)\widehat{h}_k-
2\widehat{h}_k^{\top}\Big(\sum_{t\in C_k}x_tu_t\Big)\right].
\end{equation}
We write $L_k$ for the length of the interval~$C_k$ and define
$$
\xi_T = \sup\limits_{r-s\ge T\delta_T}\Big\|\frac{1}{\sqrt{r-s}}\sum_{t=s}^{r-1} x_tu_t\Big\|; \qquad \qquad \nu_T = \sup\limits_{0<r-s< T\delta_T}\sum_{t=s}^{r-1}u_t^2. 
$$

When $L_k\ge T\delta_T$, we have
\begin{equation}
\label{q.form.lb}
\widehat{h}_k^{\top}\Big(\sum_{t\in C_k}x_tx_t^{\top}\Big)\widehat{h}_k\ge \underline{c}_{xx} L_k\|\widehat{h}_k\|^2
\end{equation}
by Lemma~\ref{lem.suppl}(ii). We also have $\widehat{h}_k^{\top}(\sum_{t\in C_k}x_tu_t)\le\sqrt{L_k}\|\widehat{h}_k\|\xi_T$, which implies
\begin{equation}
\label{crosspd.ub1}
\widehat{h}_k^{\top}\Big(\sum_{t\in C_k}x_tu_t\Big)\le c_1L_k\|\widehat{h}_k\|^2+c_2\xi^2_T
\end{equation}
for some constants~$c_1, c_2$, where we can choose an arbitrarily small~$c_1$  by increasing~$c_2$. Thus, when $L_k\ge T\delta_T$ , the contribution to~(\ref{Q.eq.rwr}) of the $k$-th summand satisfies the following inequality:
\begin{equation}
\label{Q.eq.rwr.ge.term}
\widehat{h}_k^{\top}\Big(\sum_{t\in C_k}x_tx_t^{\top}\Big)\widehat{h}_k-
2\widehat{h}_k^{\top}\Big(\sum_{t\in C_k}x_tu_t\Big)\ge (\underline{c}_{xx} -c_1)L_k\|\widehat{h}_k\|^2-c_2\xi^2_T.
\end{equation}

When $L_k<T\delta_T$, we have $\widehat{h}_k^{\top}(\sum_{t\in C_k}x_tu_t)\le \big[\widehat{h}_k^{\top}(\sum_{t\in C_k}x_tx_t^{\top})\widehat{h}_k\big]^{1/2} \big[\sum_{t\in C_k}u_t^2\big]^{1/2}$, and hence
\begin{equation}
\label{crosspd.ub2}
\widehat{h}_k^{\top}\Big(\sum_{t\in C_k}x_tu_t\Big)\le c_3\widehat{h}_k^{\top}\Big(\sum_{t\in C_k}x_tx_t^{\top}\Big)\widehat{h}_k + c_4\nu_T, 
\end{equation}
where we can again choose an arbitrarily small~$c_3$ by increasing~$c_4$. Thus, when $L_k< T\delta_T$ , the contribution to~(\ref{Q.eq.rwr}) of the $k$-th summand satisfies the following inequality:
\begin{equation}
\label{Q.eq.rwr.less.term}
\widehat{h}_k^{\top}\Big(\sum_{t\in C_k}x_tx_t^{\top}\Big)\widehat{h}_k-
2\widehat{h}_k^{\top}\Big(\sum_{t\in C_k}x_tu_t\Big)\ge (1-c_3)\widehat{h}_k^{\top}\Big(\sum_{t\in C_k}x_tx_t^{\top}\Big)\widehat{h}_k - c_4\nu_T.
\end{equation}

Combining inequalities (\ref{Q.eq.rwr}), (\ref{Q.eq.rwr.ge.term}), (\ref{Q.eq.rwr.less.term}), and using $c_1\le\underline{c}_{xx}$ together with $c_3\le 1$, we derive
\begin{equation*}
Q(\widehat{\bbeta})-Q({\bbeta}^*)\ge -(c_2+c_4)\sum_k (\xi_T^2+\nu_T).
\end{equation*}
Because the number of terms in the sum $\sum_k$, i.e. the total number of intervals $C_k$ is at most $\widehat{m}+m^*+1$, we deduce that
\begin{equation*}
Q(\widehat{\bbeta})-Q({\bbeta}^*)\ge -c_5(\xi_T^2+\nu_T)(\widehat{m}+m^*).
\end{equation*}
Noting that $Q(\widehat{\bbeta})+\lambda\widehat{m}\le Q({\bbeta}^*)+\lambda m^*$, we conclude that 
\begin{equation}
\label{Q.lam.ineq}
\lambda\widehat{m}\le \lambda m^*+O_p\Big([\xi_T^2+\nu_T][\widehat{m}+m^*]\Big).
\end{equation}
Observe that $\xi_T^2=(\log T)^{c_{\delta}}$ and $\nu_T=O_p(T\delta_T)$ by Lemma~\ref{lem.suppl}(iii) and Lemma~\ref{lem.suppl}(iv), respectively. Hence, $\xi_T^2+\nu_T= O_p(T\delta_T)$ by Assumption A2(ii), and thus $(\xi_T^2+\nu_T)(m^*+1)=o_p(\lambda)$ by the assumed lower bound on~$\lambda$, which, in turn, implies $[\xi_T^2+\nu_T][\widehat{m}+m^*]=o_p(\lambda|\widehat{m}-m^*|+\lambda)$. Consequently, we can rewrite inequality~(\ref{Q.lam.ineq}) as
$$
\widehat{m} \le  m^*+o_p(|\widehat{m}-m^*|)+o_p(1),
$$
and hence $\widehat{m}\le m^*$ with probability tending to one.

We will now argue by contradiction to establish that, with probability tending to one, within $I_{\text{min}}/5$ of each true breakpoint lies an estimated breakpoint. Suppose that this is false, and hence, with positive non-vanishing probability, there exists a (randomly selected) true breakpoint~$T^*_{\tilde{k}}$, such that no estimated breakpoints are within $I_{\text{min}}/5$ of~$T^*_{\tilde{k}}$.

We define $\tilde{C}_{-}=\{1,...,T\}\cap(T^*_{\tilde{k}}-I_{\text{min}}/5\,,\,T^*_{\tilde{k}})$ and $\tilde{C}_{+}=\{1,...,T\}\cap[T^*\,,\,T^*_{\tilde{k}}+I_{\text{min}}/5)$. 
We set $\tilde{\bbeta}$ equal $\widehat{\bbeta}$ for all~$t$ except the ones falling in~$\tilde{C}$, where we set $\tilde{\bbeta}_t=\bbeta^*_t$. 
Note that
\begin{equation}
Q(\widehat{\bbeta})-Q({\tilde{\bbeta}})=\sum_{t\in \tilde{C}_{-}\cup \tilde{C}_{+}}\Big[(\tilde{\bbeta}_t-\beta^*_t)^{\top}(x_tx_t^{\top})(\tilde{\bbeta}_t-\bbeta^*_t)-
2(\tilde{\bbeta}_t-\bbeta^*_t)^{\top}x_tu_t\Big].
\end{equation}
Observing that the vector of estimated regression coefficients stays constant in the interval $\tilde{C}_{-}\cup \tilde{C}_{+}$, we denote this vector by~$\widehat{\gamma}$. We write~$\gamma_1^*$ and~$\gamma_2^*$ for the true regression coefficient vectors corresponding to the time points in~$\tilde{C}_{-}$ and~$\tilde{C}_{+}$, respectively. Assumptions A2(ii) and A3(ii) imply that $I_{\text{min}}\rightarrow\infty$ as $T\rightarrow\infty$. Hence, the lengths of the intervals corresponding to~$\tilde{C}_{-}$ and~$\tilde{C}_{+}$ are approximately~$I_{\text{min}}/5$. Assumption A3(ii) further implies that $T\delta_T=o(I_{\text{min}})$. Thus, applying inequalities (\ref{q.form.lb})-(\ref{crosspd.ub1}) with $L_k\ge I_{\text{min}}/6$ and $c_1<\underline{c}_{xx}$, we derive
\begin{eqnarray*}
Q(\widehat{\bbeta})-Q(\tilde{\bbeta}) &\ge& c_6 I_{\text{min}}\big(\|\widehat{\gamma}-\gamma^*_1\|^2 + \|\widehat{\gamma}-\gamma^*_2\|^2 \big)- c_2\xi_T^2\\
&\ge& (c_6/2) I_{\text{min}} J_{\text{min}}^2  - c_2\xi_T^2,
\end{eqnarray*}
for some positive constants $c_6$ and $c_2$. Let $\tilde{m}$ be the number of breakpoints corresponding to~$\tilde{\bbeta}$, and note that $\tilde{m}\le \widehat{m}+2$. Because $Q(\widehat{\bbeta})+\lambda\widehat{m}\le Q(\tilde{\bbeta})+\lambda \tilde{m}$, we can then deduce that inequality
$$
\lambda\ge c_7I_{\text{min}}J_{\text{min}}^2-c_8\xi_T^2
$$
holds with positive non-vanishing probability. 
Using the following bounds we established earlier: $\xi_T^2 = O_p([\log T]^{c_{\delta}})$ and $T\delta_T=o(I_{\text{min}})$, together with the Assumption A3(i) on $J_{\text{min}}^2$, we derive that $\xi_T^2=o(I_{\text{min}}J_{\text{min}}^2)$, and hence
$\lambda\ge c_9I_{\text{min}}J_{\text{min}}^2$  with positive non-vanishing probability. However, this last inequality constitutes a contradiction with the assumption $\lambda/\big[J^2_{\text{min}}I_{\text{min}}\big]\rightarrow 0$ imposed in the statement of Theorem~\ref{asympt.theorem}.


Thus, we have established that the following two statements hold with probability tending to one: (a) $\widehat{m}\le m^*$; and (b) within $I_{\text{min}}/5$ of each true breakpoint $T^*_1,...,T^*_{m^*}$ lies an estimated breakpoint. It follows directly that $\widehat{m}=m^*$, which completes the proof of claim~1.

\medskip

\noindent\textbf{Claim 2: }$\mathbf{P}\Big(\textbf{max}_{1\le j\le m^*}\big|\widehat{T}_j-T^*_j\big|\le T{\delta}_T\Big)\rightarrow 1$.
We restrict our attention to the set of probability tending to one where statements (a) and (b) in the above paragraph are satisfied. Because $\widehat{m}=m^*$ and $|\widehat{T}_j-T^*_j|\le I_{\text{min}}/5$ for each $j=1,...,m^*$, the length of the interval where the estimated coefficient vector is $\widehat{\bbeta}_j$ while the true coefficient vector is $\bbeta^*_j$ is at least $3I_{\text{min}}/5$. Recall the earlier established bound $T\delta_T=o(I_{\text{min}})$. Again applying inequalities (\ref{Q.eq.rwr}), (\ref{Q.eq.rwr.ge.term}), (\ref{crosspd.ub2}), with $c_1<\underline{c}_{xx}$ and $c_3\le 1$, we derive
\begin{equation*}
Q(\widehat{\bbeta})-Q(\bbeta^*) \ge c_{10} I_{\text{min}}\Big(\sum_{j=1}^{m^*}\|\widehat{\bbeta}_j-\bbeta^*_j\|^2\Big)- c_{11}\big(\xi_T^2+\nu_T\big)m^*
\end{equation*}
for some positive constants $c_{10}$ and $c_{11}$. Recall that $\xi_T^2+\nu_T=O_p(T\delta_T)$. Thus, taking into account $Q(\widehat{\bbeta})\le Q(\bbeta^*)$, we can conclude that
$$
\sum_{j=1}^{m^*}\|\widehat{\bbeta}_j-\bbeta^*_j\|^2=O_p\left(\frac{m^*T\delta_T}{I_{\text{min}}}\right).
$$
By the assumptions imposed on~$\lambda$, the right hand side of the above inequality is $o_p(J^2_{\text{min}})$. Consequently, and because $\min_{j\ne k}\|\bbeta^*_j-\bbeta^*_k\|^2\ge J^2_{\text{min}}$, we arrive at $\min_{j\ne k}\|\widehat{\bbeta}_j-\bbeta^*_k\|^2\ge J^2_{\text{min}}/5$.

We define $L_T=\max_{1\le j\le m^*}|\widehat{T}_j-T^*_j|$ and argue by contradiction to establish $L_T\le T\delta_T$ with probability tending to one. Suppose that this claim is false, i.e., $L_T> T\delta_T$ with positive non-vanishing probability. Let~$T^*_{\tilde{k}}$ denote the true breakpoint (randomly selected) where $\max_{1\le j\le m^*}|\widehat{T}_j-T^*_j|$ is achieved, that is, $L_T=|\widehat{T}_{\tilde{k}}-T^*_{\tilde{k}}|$.  
For concreteness, suppose that $\widehat{T}_{\tilde{k}}>T^*_{\tilde{k}}$. The complimentary case can be handled by nearly identical arguments with minor notational modifications. To simplify the presentation, we will write $\widehat{\gamma}_1$ for the estimated regression coefficient vector in the interval $(\widehat{T}_{\tilde{k}-1},\widehat{T}_{\tilde{k}})$ and write $\widehat{\gamma}_2$ for the estimated regression coefficient vector in the interval $(\widehat{T}_{\tilde{k}},\widehat{T}_{\tilde{k}+1})$. Similarly, we use~$\gamma^*_1$ for the true regression coefficients in $(T^*_{\tilde{k}-1},T^*_{\tilde{k}})$ and~$\gamma^*_2$ for the ones in $(T^*_{\tilde{k}-1},T^*_{\tilde{k}+1})$.
Let $\tilde{\bbeta}$ equal $\widehat{\bbeta}$ for all~$t$ except the ones in the interval $(T^*_{\tilde{k}},\widehat{T}_{\tilde{k}})$, where we set $\tilde{\beta}_t=\widehat{\gamma}_2$. Note that the number of breakpoints corresponding to~$\tilde{\bbeta}$ is still~$m^*$, and hence $Q(\widehat{\bbeta})\le Q(\tilde{\bbeta})$. Consequently, applying inequalities (\ref{q.form.lb})-(\ref{crosspd.ub2}) with a sufficiently small~$c_1$ once again, and collecting the terms, we deduce that inequality
\begin{equation}
\label{LT.equation}
  L_T\|\widehat{\gamma}_1 - \gamma_2^*\|^2 = O_p\left([\log T]^{c_{\delta}}+L_T\|\widehat{\gamma}_2 - \gamma_2^*\|^2 \right),  
\end{equation}
holds with positive non-vanishing probability.  We showed earlier that $\|\widehat{\gamma}_2-\gamma^*_2\|^2=o_p(J^2_{\text{min}})$ and $\|\widehat{\gamma}_1-\gamma^*_2\|^2\ge J^2_{\text{min}}/5$. Hence, inequality~(\ref{LT.equation}) gives 
$$
L_T= O_p\left(\frac{[\log T]^{c_{\delta}}}{J^2_{\text{min}}}\right)+o_p(L_T),
$$
which implies $L_T= O_p\Big([\log T]^{c_{\delta}}J^{-2}_{\text{min}}\Big)$.
By Assumption A3(i), we thus have $L_T=o_p(T\delta_T)$,
which contradicts our starting assumption that $L_T> T\delta_T$ with non-vanishing positive probability.

\medskip

\noindent\textbf{Claim 3: }${\widehat{\balpha}_j-\balpha^*_j=O_p\big([I_j^*]^{-1/2}\big)}$. It is only left to establish the stated rate of convergence for the regression coefficients $\widehat{\balpha}_j$. This result follows directly from Theorem 3.1(ii) in \cite{qian_su_2016} after setting the $\ell_1$ penalty weight (their~$\lambda$ parameter) to zero and recalling that $\delta_T=O\big(I_{\textrm{min}}^{1/2}/T\big)$. While the result in  \cite{qian_su_2016} has an additional assumption $m^*=O(\log T)$, an analysis of their proof reveals that this assumption is not required as long as $\max_{1\le j\le m^*}\big|\widehat{T}_j-T^*_j\big|\le T\delta_T$ with probability tending to one, which is a property that we established in the previous paragraph.  
\end{proof}

\subsection*{Appendix D: Proof of Theorem~\ref{CLT}}
\begin{proof}
Let $\widehat{\balpha}^*=\big(\mathbb{X}^{\top}\mathbb{X}\big)^{-1}\mathbb{X}^{\top}Y$. In the proof of their Theorem~3.6, \cite{qian_su_2016} show that~$\widehat{\balpha}^*$ has the asymptotic distribution specified in the statement of Theorem~\ref{CLT}, i.e., ${D}(\widehat{\balpha}^*-\balpha^*)\xrightarrow[]{d}N\big(\mathbf{0},\Psi^{-1}\Phi\Psi^{-1}\big)$. Consequently, to complete the proof, it is sufficient to establish
\begin{equation}
\label{prof.stoch.bnd}
   \widehat{D}(\widehat{\balpha}_{m^*}-\balpha^*)-D(\widehat{\balpha}^*-\balpha^*)=o_p(1). 
\end{equation} 
The above stochastic bound is derived by \cite{qian_su_2016} for their post-LASSO estimator in the proof of their Theorem~3.6. However, an analysis of the proof reveals that, under our imposed assumptions,  bound~(\ref{prof.stoch.bnd}) holds for any estimator $\tilde{\balpha}$ of the form $\tilde{\balpha}=\big(\tilde{\mathbb{X}}^{\top}\tilde{\mathbb{X}}\big)^{-1}\tilde{\mathbb{X}}^{\top}Y$, where $\tilde{\mathbb{X}}=\text{diag}(\tilde{\mathbb{X}}_1,...,\tilde{\mathbb{X}}_{m^*+1})$ and $\tilde{\mathbb{X}}_j=(x_{\tilde{T}_{j-1}},...,x^{\top}_{\tilde{T}_{j}-1})$, such that  $P(\max_{1\le j\le m^*}\big|\tilde{T}_j-T^*_j\big|\le T\delta_T)\rightarrow 1$ as $T\rightarrow\infty$. By Theorem~\ref{asympt.theorem}, this condition is satisfied for our estimator~$\widehat{\balpha}_{m^*}$.
\end{proof}





\subsection*{Appendix E: Simulations for the case of no breaks}

The Monte Carlo simulations reported in this section are based on the data generating processes (DGP) similar to those used by \cite{qian_su_2016}\footnote{\citet{qian_su_2016} added a constant to the DGP. We omit it to be consistent with the DGPs in Sections 4.2 and 4.3.}. We simulate the data as follows:
\begin{equation}
    y_t=x_t+u_t,
\end{equation}

\begin{table}[!htbp]
\centering
\footnotesize
\caption{Correct detections: zero breaks.}\label{tab1_0breaks}
\begin{tabular}{cccccccc} \toprule
 &$\sigma_u$&$T$&MIO&GFL&BIC&LWZ&SEQ\\\hline
DGP-1&0.5&100&96.2&97.8&96.8&100.0&86.8\\
 &&200&99.8&99.8&98.6&100.0&92.0\\
 &&500&100.0&100.0&98.8&100.0&94.8\\
 &1&100&96.6&97.8&96.8&100.0&86.8\\
 &&200&99.8&99.8&98.6&100.0&92.0\\
 &&500&100.0&100.0&98.8&100.0&94.8\\
 &1.5&100&96.6&97.8&96.8&100.0&86.8\\
 &&200&99.8&99.8&98.6&100.0&92.0\\
 &&500&100.0&100.0&98.8&100.0&94.8\\\hline
DGP-2&0.5&100&95.6&97.6&97.2&99.8&86.6\\
 &&200&99.6&100.0&98.2&100.0&90.6\\
 &&500&100.0&100.0&98.4&100.0&93.0\\
 &1&100&95.6&97.6&97.2&99.8&86.6\\
 &&200&99.6&100.0&98.2&100.0&90.6\\
 &&500&100.0&100.0&98.4&100.0&93.0\\
 &1.5&100&95.8&97.6&97.2&99.8&86.6\\
 &&200&99.6&100.0&98.2&100.0&90.6\\
 &&500&100.0&100.0&98.4&100.0&93.0\\\hline
DGP-3&0.5&100&92.6&97.6&97.0&100.0&86.4\\
 &&200&97.6&99.2&97.2&100.0&90.8\\
 &&500&100.0&100.0&99.4&100.0&92.6\\
 &1&100&93.2&97.6&97.0&100.0&86.4\\
 &&200&97.6&99.2&97.2&100.0&90.8\\
 &&500&100.0&100.0&99.4&100.0&92.6\\
 &1.5&100&93.6&97.6&97.0&100.0&86.4\\
 &&200&97.6&99.2&97.2&100.0&90.8\\
 &&500&100.0&100.0&99.4&100.0&92.6\\
\hline

DGP-4&0.5&100&100.0&100.0&99.6&100.0&93.6\\
 &&200&96.2&97.8&97.4&99.8&89.8\\
 &&500&99.6&99.8&97.0&100.0&90.8\\
 &1&100&99.8&100.0&99.4&100.0&95.0\\
 &&200&88.4&92.6&93.2&99.2&89.8\\
 &&500&93.4&93.2&93.2&99.4&92.2\\
 &1.5&100&98.6&94.8&93.6&99.6&95.4\\
 &&200&88.6&95.8&95.8&99.6&89.4\\
 &&500&96.4&98.8&95.4&100.0&90.6\\\hline
DGP-5&$\sigma_2=0.2$&100&100.0&100.0&99.0&100.0&93.6\\
 &&200&96.6&98.2&97.6&99.8&87.2\\
 &&500&99.4&99.8&97.4&100.0&89.6\\
 &$\sigma_2=0.3$&100&100.0&100.0&100.0&100.0&95.2\\
 &&200&90.0&95.6&95.8&99.8&88.4\\
 &&500&96.8&99.0&96.4&99.8&90.4\\
 &$\sigma_2=0.5$&100&100.0&100.0&99.0&100.0&96.6\\
 &&200&97.2&98.0&97.2&100.0&81.2\\
 &&500&99.8&99.8&98.0&100.0&88.6\\\hline
DGP-6&$a=0.2$&100&100.0&100.0&99.4&100.0&91.4\\
 &&200&96.8&98.2&97.4&99.6&83.0\\
 &&500&99.4&99.6&98.4&100.0&90.2\\
 &$a=0.5$&100&100.0&100.0&99.2&100.0&91.6\\
 &&200&96.2&97.6&96.0&99.6&81.4\\
 &&500&99.8&100.0&97.6&100.0&88.6\\
 &$a=0.9$&100&100.0&100.0&98.2&100.0&88.6\\
 &&200&99.8&100.0&97.6&100.0&88.6\\
 & &500&100.0&100.0&98.2&100.0&88.6\\
\hline
   \bottomrule
\end{tabular}
\end{table}

where
\begin{enumerate}
    \item $x_t\sim i.i.d. N(0,1)$, $u_t\sim i.i.d. N(0,\sigma_u^2)$
    \item $x_t=0.5x_{t-1}+\eta_t$, $\eta_t\sim i.i.d. N(0,0.75)$, $u_t\sim i.i.d. N(0,\sigma_u^2)$
    \item $x_t\sim i.i.d. N(0,1)$, $u_t=\sigma_uv_t$, $v_t=0.5v_{t-1}+\varepsilon_t$, $\varepsilon_t\sim i.i.d. N(0,1)$
    \item $x_t=0.5x_{t-1}+\eta_t$, $\eta_t\sim i.i.d. N(0,0.75)$, $u_t=\sigma_u\sqrt{h_t}\varepsilon_t$, $h_t=0.05+0.05u_{t-1}^2+0.9h_{t-1}$, $\varepsilon_t\sim i.i.d. N(0,1)$
    \item $x_t=0.5x_{t-1}+\eta_t$, $\eta_t\sim i.i.d. N(0,0.75)$, $u_t\sim i.i.d. N(0,\sigma_1^2)$ for $t\in\{1,2,\dots,T/2\}$ and $u_t\sim i.i.d. N(0,\sigma_2^2)$ for $t\in\{T/2,T/2+1,\dots,T\}$ 
    \item $y_t=\alpha y_{t-1}+\varepsilon_t$, $x_t=y_{t-1}$, $\varepsilon_t\sim i.i.d. N(0,1-\alpha^2)$
\end{enumerate}

The six DGP variations assume diverse types of serial correlation, conditional heteroskedasticity and causality of variables. The true values of parameters are $\sigma_u\in\{0.5,1,1.5\}$, $\sigma_1=0.1$, $\sigma_2\in\{0.2,0.3,0.5\}$, $\alpha\in\{0.2,0.5,0.9\}$. 

Table \ref{tab1_0breaks} reports the fraction of replications that correctly detected 0 breaks (in percentages). 
We observe that for all GDPs, MIO and GFL produce similar percentages of correct detection of no breaks. BIC and SEQ perform worse uniformly, while LWZ remarkably gives almost 100\% correct detection of no breaks in all GDPs. For DGP 1-3, the fraction of correct detections increases as the sample size increases suggesting that the estimators are consistent. For DGP 4-6, there are some deviations from consistency which we attribute to finite samples and the more complex dynamics, i.e., heteroskedastic errors, a break in error variance and an endogenous regressor.

\end{document}